\documentclass[aps,prl,twocolumn,superscriptaddress,longbibliography,notitlepage,10pt]{revtex4-1}
\usepackage{amsmath,amsfonts,amssymb,color,epsfig,graphics,graphicx,latexsym,theorem,url,multirow}
\usepackage[colorlinks,colorlinks,citecolor=blue,linkcolor=blue,urlcolor=blue]{hyperref}
\newtheorem{definition}{Definition}
\newtheorem{theorem}[definition]{Theorem}
\usepackage[utf8]{inputenc}
\usepackage[T1]{fontenc}
\usepackage{courier}
\usepackage{listings}
\def\tightlist{}
\usepackage{fixltx2e} 

\usepackage{natbib}

\usepackage{graphicx}
\graphicspath{{}}

\begin{document}
\begin{abstract}
Representation by neural networks, in particular by restricted Boltzmann machines (RBM), has provided a powerful computational tool to solve quantum many-body problems. An important open question is how to characterize which class of quantum states can be efficiently represented with RBMs. Here, we show that RBMs can efficiently represent a wide class of many-body entangled states with rich exotic topological orders. This includes: (1) ground states of double semion and twisted quantum double models with intrinsic topological orders; (2) states of the AKLT model and two-dimensional CZX model with symmetry protected topological orders; (3) states of Haah code model with fracton topological order; (4) (generalized) stabilizer states and hypergraph states that are important for quantum information protocols. One twisted quantum double model state considered here harbors non-abelian anyon excitations. Our result shows that it is possible to study a variety of quantum models with exotic topological orders and rich physics using the RBM computational toolbox.
\end{abstract}

\title{Efficient Representation of Topologically Ordered States with Restricted Boltzmann Machines}

\author{Sirui Lu}
\affiliation{Department of Physics, Tsinghua University, Beijing 100084, China}
\affiliation{Center for Quantum Information, Institute for Interdisciplinary Information Sciences, Tsinghua University, Beijing 100084, China}
\author{Xun Gao}
\email{gaoxungx@gmail.com}
\affiliation{Center for Quantum Information, Institute for Interdisciplinary Information Sciences, Tsinghua University, Beijing 100084, China}
\affiliation{Department of Physics, Harvard University, Cambridge, MA 02138, USA}
\author{L.-M. Duan}
\email{lmduan@tsinghua.edu.cn}
\affiliation{Center for Quantum Information, Institute for Interdisciplinary Information Sciences, Tsinghua University, Beijing 100084, China}

\date{\today}

\maketitle

\newcommand{\plusnamesingular}{}
\newcommand{\starnamesingular}{}
\newcommand{\xrefname}[1]{\protect\renewcommand{\plusnamesingular}{#1}}
\newcommand{\Xrefname}[1]{\protect\renewcommand{\starnamesingular}{#1}}
\providecommand{\cref}{\plusnamesingular~\ref}
\providecommand{\Cref}{\starnamesingular~\ref}
\providecommand{\crefformat}[2]{}
\providecommand{\Crefformat}[2]{}

\crefformat{figure}{Fig.~#2#1#3}
\Crefformat{figure}{Figure~#2#1#3}
\crefformat{equation}{Eq.~#2#1#3}
\Crefformat{equation}{Equation~#2#1#3}

\emph{Introduction.}---Deep learning has become a powerful tool with wide applications \citep{LeCun:2015dt, goodfellow2016deep}. Recently, deep learning methods have attracted considerable attention in quantum physics \citep{Biamonte:2017ica, ciliberto2018quantum}, especially for attacking quantum many-body problems. The difficulty of quantum many-body problems mainly originates from the exponential growth of the Hilbert space dimension. To overcome this exponential difficulty, researchers traditionally use tensor network methods \citep{schollwock2011density, verstraete2008matrix, schuch2008simulation} and Quantum Monte Carlo (QMC) simulation \citep{ceperley1986quantum}. However, QMC methods suffer from the sign problem \citep{loh1990sign}; Tensor network methods have difficulty to deal with high dimensional systems \citep{schuch2007computational} or systems with massive entanglement \citep{verstraete2006criticality}. These issues call for mew method.

Being one of the fundamental building block of deep learning, the neural network has been recently employed as a compact representation of quantum many-body states \citep{Carleo602, deng2017machine, clark2018unifying, Gao:2017uk, huang2017neural, deng2017quantum, cai2018approximating, chen2018equivalence, glasser2018neural, jia2018efficient, carleo2018constructing}. Many variants of neural networks have been investigated numerically or theoretically, such as the restricted Boltzmann machines (RBM) \citep{Carleo602, deng2017machine, Gao:2017uk, jia2018efficient}, the deep Boltzmann machine (DBM) \citep{Gao:2017uk, carleo2018constructing, huang2017neural}, and the feed-forward Neural Network (FNN)\citep{cai2018approximating}. The RBM ansatz have also been investigated for quantum information protocols \citep{Torlai:2018wna, jonsson2018neural, deng2018machine}. We focus here on the RBM states which work efficiently during variational optimization although the representational power of which is somewhat limited. An important open issue is how to characterize the class of quantum many-body states that can be represented by the RBMs.

In the past decades, the studies of topological order
\citep{zeng2015quantum, wen2004quantum}, which are beyond the framework of Landau's symmetry breaking paradigm \citep{landau1980statistical}, have attracted tremendous attention. There are several types of topological ordered states: the intrinsic topological ordered states feature unliftable ground state degeneracy through local perturbations; the symmetry protected topological (SPT) ordered states with a given symmetry cannot be smoothly deformed into each other without a phase transition if the deformation preserves the symmetry; the fracton topological ordered states harbor point excitations that are immobile in the three-dimensional space, i.e., fractons. While several studies have shown that the RBM can capture simple many-body states such as graph/cluster states \citep{Gao:2017uk, deng2017machine} and toric/surface code states \citep{Gao:2017uk, deng2017machine, jia2018efficient}, no single study exists which represents other more exotic topological states in the condensed matter physics \citep{wen2004quantum, zeng2015quantum, Chen1604, RevModPhys.88.035005}.

In this paper, we use tools from quantum information to construct the RBM representations for other notable many-body states, focusing on different topologically ordered states. Many of exotic condensed matter topological states can be described by powerful quantum information tools: (i) the hypergraph state formalism which generalizes the graph-state formalism; (ii) the stabilizer formalism \citep{gottesman1997stabilizer} which describes most of the quantum error correction code; (iii) the XS-stabilizer formalism \citep{ni2015non} which generalizes stabilizer formalism. These formalisms themselves are vital for quantum error correction \citep{gottesman1997stabilizer}, classical simulation of quantum circuits \citep{gottesman1998heisenberg} and Bell's nonlocality \citep{gachechiladze2016extreme, guhne2014entanglement, bell2001einstein, RevModPhys.86.419, scarani2005nonlocality}. We prove these states of (i-iii) can be represented by the RBM efficiently based on the properties of their wave functions. We also propose a unary representation to generalize RBM state to higher spin systems.

These tools from quantum information provide recipes for constructing RBM representation within their formalism. The stabilizer formalism describes many fracton models \citep{haah2011local, yoshida2013exotic, ma2017fracton, vijay2015new, vijay2016fracton, chamon2005quantum} such as Haah's code \citep{haah2011local}. Concerning the intrinsic topological order, we show tbe RBM states can capture double semion of string-net model \citep{levin2005string, gu2009tensor} and many twisted quantum double model \citep{kitaev2003fault, hu2013twisted, beigi2011quantum} using their XS-stabilizer description. For symmetry protected topological orders, we give exact constructions of the AKLT model \citep{affleck1987rigorous, affleck1988valence} with the unary representation. We also consider RBM representation for other SPT models such as the two-dimensional (2D) CZX model \citep{chen2011two}. Our exact representation results provide insights and a powerful tool for future studies of quantum topological phase transitions and quantum information protocols.

\emph{RBM state.---} We first recall the definition of RBM state and describe notations. In the computational basis, a quantum wave function of \(n\)-qubit can be expressed as \(|\Psi\rangle=\sum_{\mathbf{v}}\Psi(\mathbf{v})|\mathbf{v}\rangle\) with \(\mathbf{v}\equiv (v_1,\dots,v_n)\), where the \(\Psi(\mathbf{v})\) is a complex function of \(n\) binary variables \(v_i\). We use \(\{0, 1\}\) valued vertices instead of \(\{-1, 1\}\) valued vertices for convenience. In the case of RBM, \(\Psi(\mathbf{v})=\sum_{\mathbf{h}}e^{W(\mathbf{v,h})}\), where the weight \(W(\mathbf{v,h})=\sum_{i,j} W_{ij}v_i h_j+\sum_i a_i v_i+\sum_j b_j h_j\) is a complex quadratic function of binary variables. While a Boltzmann machine allows arbitrary intra-layer connection, in RBM the visible neurons \(\mathbf{v}\) only connect to hidden neurons \(\mathbf{h}\). Let the number of visible neurons be \(n\) and the number of hidden neurons be \(m\). We say the representation is efficient if \(m=\text{poly}(n)\). The whole wave function writes
\begin{equation}
\Psi_{\text{RBM}}(\mathbf{v})=e^{\sum_{i=1}^n a_i v_i}\prod_{j=1}^m (1+\exp\left(\theta_j\right))
\label{eq:RBM}\end{equation}
with effective angles \(\theta_j=b_j+\sum_{k=1}^n W_{kj} v_k\).

\begin{figure}
\centering
\includegraphics[width=0.40000\textwidth]{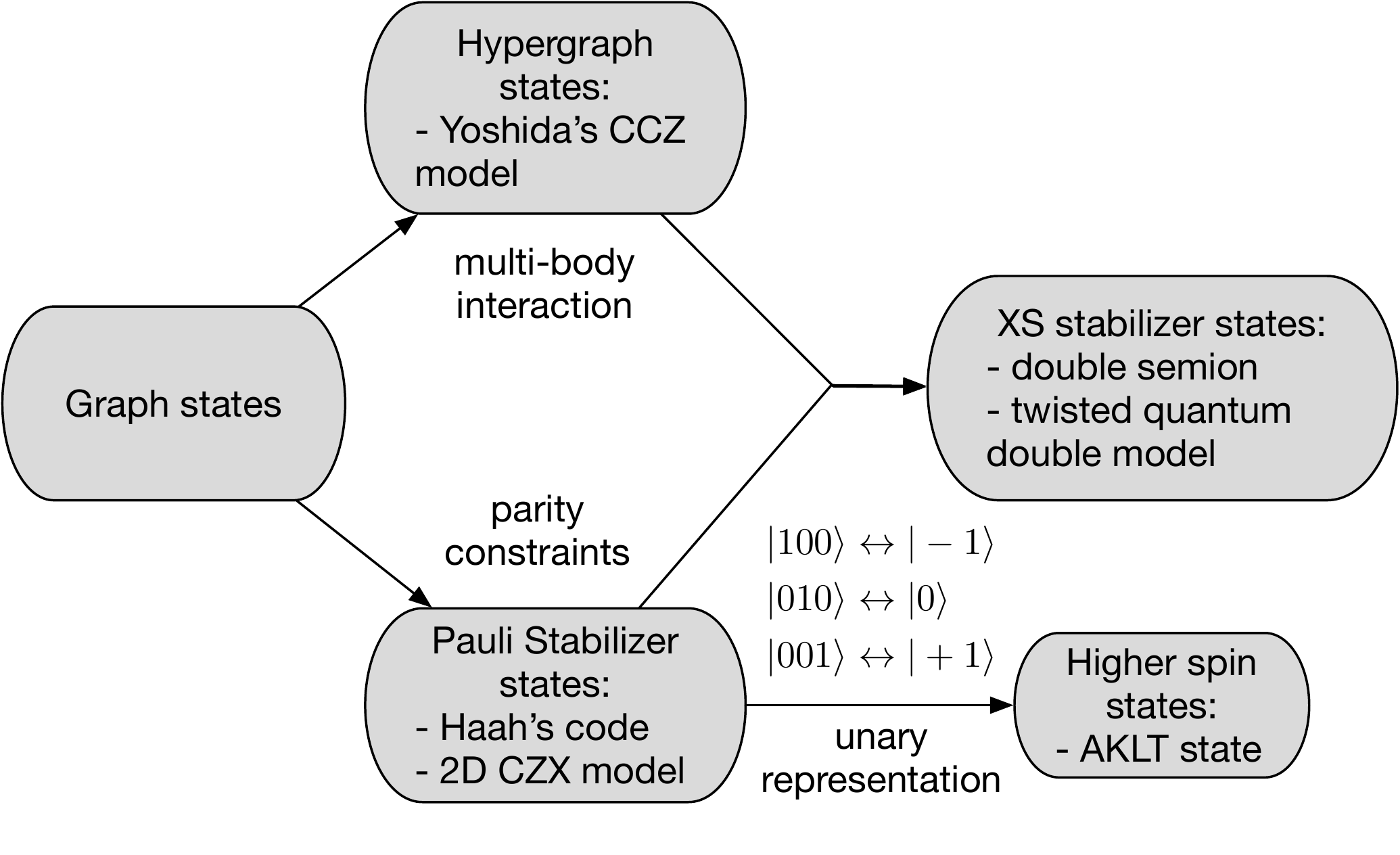}
\caption{Different generalizations of graph states . (a) Hypergraph states generalize graph states by introducing 3-body correlation factors; (b) Stabilizer states generalize graph states by additional parity constraints on qubits; (c) XS-stabilizer states combine the parity constraints from Pauli stabilizers and 3-body correlation factors from hypergraph states. Many condensed matter topological models fall into these quantum information formalisms, thus can be represented efficiently by RBM. We will encounter them later. Combining with the unary representation, stabilizer states with added unary constraints can describe the AKLT state. Thus the AKLT state can be efficiently represented by the RBM.\label{fig:paradigm}}
\end{figure}

We depict this paper's roadmap in \xrefname{Fig.}\cref{fig:paradigm}. RBM states \citep{hein2006entanglement} have been shown to represent graph states efficiently \citep{Gao:2017uk, deng2017machine} (recall the wave function of graph states takes the form \(\Psi(v_1,\cdots,v_n)=\prod_{\{ i,j\} }(-1)^{v_i v_j}\) (up to a normalization factor), where \(\{ i,j\}\) denotes an edge linking the \(i\)-th and \(j\)-th qubits represented by visible neurons \(v_i\), \(v_j\)). There are various ways to generalize graph states. Hypergraph states \citep{rossi2013quantum} generalize graph states by introducing more than two body correlation factors such as \((-1)^{v_1v_2v_3}\). Stabilizer states generalize graph states through additional local Clifford operations \citep{dehaene2003clifford, van2004graphical, grassl2002graphs}, which impose parity constraints and extra phases. XS-stabilizer states \citep{ni2015non} combines 3-body correlation factors from hypergraph state and parity constraints from stabilizer states. A parity constraint: \((v_1+v_2+\cdots+v_k)\mod 2=0\) can be realized by a hidden neuron that connects to each of these visible neurons \(v_1,v_2,\cdots,v_k\) with weight function \(W(v,h)=i\pi vh-(\ln 2)/4\) \citep{affine}. Next we derive the RBM representation of multi-body correlation factors (hypergraph states) and the unary representation, making all the states in \xrefname{Fig.}\cref{fig:paradigm} can be represented by the RBM.

\begin{figure}
\centering
\includegraphics[width=0.40000\textwidth]{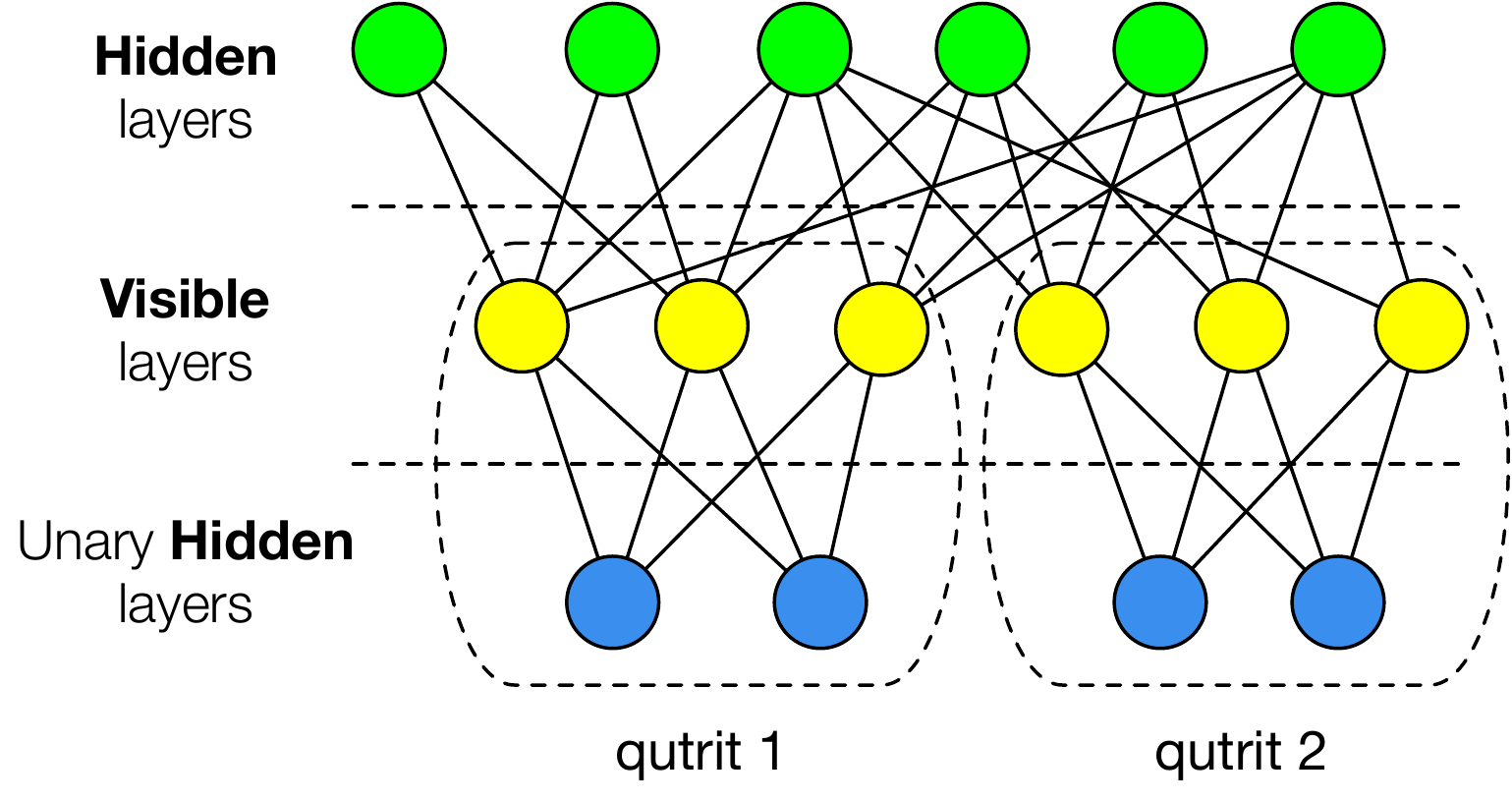}
\caption{The unary RBM representation for higher spin systems. The three left yellow neurons correspond to the qubit 1, and the three right yellow neurons correspond to the qubit 2. Two blue neurons in the unary hidden layers restrict three visible yellow neurons to be one of 100, 010 and 001 that corresponds to \(|-1\rangle\), \(|0\rangle\) and \(|+1\rangle\) in spin-1. The whole graph is still a bipartite graph.\label{fig:unary}}
\end{figure}

\emph{The unary RBM representation.---} To study higher spin systems, we propose the unary representation. The idea of unary representation is best illustrated using an example, as depicted in \xrefname{Fig.}\cref{fig:unary}. We use three neurons (qubits) to represent a spin-1: \(|100\rangle\to|-1\rangle\), \(|010\rangle\to|0\rangle\) and \(|001\rangle\to|1\rangle\) by restricting these three neurons (qubits) to the subspace spanned by \(|100\rangle,~|010\rangle\) and \(|001\rangle\). This can be done by using two hidden neurons (blue neurons in \xrefname{Fig.}\cref{fig:unary}) \citep{S}. Our unary representation is simpler than multi-value neurons or encoded binary neurons in distinguishing and decoding basis states. We will use this unary representation to represent the AKLT state.

\begin{figure}
\centering
\includegraphics[width=0.45000\textwidth]{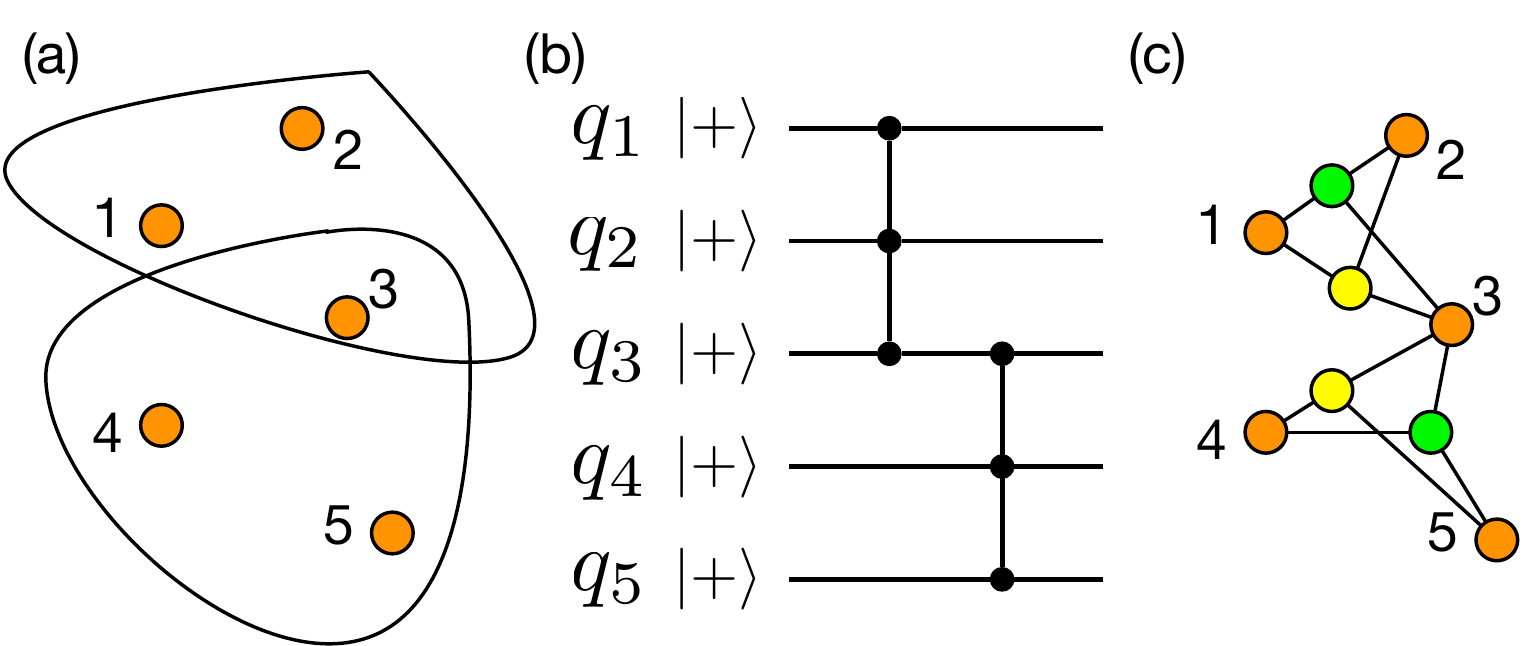}
\caption{RBM representation of hypergraph state. (a) A hypergraph. There are two 3-hyperedges that connect \({q_1,q_2,q_3}\) and \({q_3,q_4,q_5}\). (b) The encoding circuit of this hypergraph state, where \(|+\rangle = (|0\rangle+|1\rangle)/\sqrt{2}\) represents the input state for all the qubits and the connected dots denote the controlled-controlled-\(Z\) gate . (c) The corresponding RBM representation of (a). Orange neurons are visible neurons, and both green and yellow vertice are hidden neurons. The two different colors (yellow and green) represent different weight functions.\label{fig:hypergraph}}
\end{figure}

\emph{RBM representation of hypergraph states.---} Hypergraphs generalize graphs by allowing an edge can join any number of vertices. We define an edge that connects \(k\) vertices a \(k\)-hyperedge. Given a mathematical hypergraph, the wave function of its corresponding hypergraph state \citep{rossi2013quantum} takes the form:\(\Psi_{\text{hypergraph}}(\mathbf{v}) \propto\prod_{\{v_1,v_2,\dots,v_k\}\in E}(-1)^{v_1v_2\dots v_k}\). The notation \(\{v_1,v_2.\dots,v_k\}\in E\) means that these \(k\) vertices \(\{v_1,v_2.\dots,v_k\}\) are connected by a \(k\)-hyperedge. We illustrate the correspondence in \xrefname{Fig.}\cref{fig:hypergraph}. We now extend RBM representation to hypergraph states.

\begin{theorem}
Restricted Boltzmann machines can represent any hypergraph states efficiently and exactly.
\end{theorem}

In the main text, we take the graph with 3-hyperedges as an example; it will be useful later for representation of XS-stabilizer states and SPT states. Precisely, we make use of the following decomposition:
\begin{equation}
(-1)^{v_1v_2v_3}=e^{i\pi(\sum_{i=1}^3v_i)}\cdot\sum_{h_1,h_2} e^{w(\sum_{i=1}^3v_i)(h_1-h_2)+b(h_1+h_2)+c},
\label{eq:3decomp}\end{equation}
where \(w=i\), \(b=\ln(\frac{1+ \sqrt{-15}}{4})\) and \(c=\ln\frac{2}{3}-b\). Thus, the exact RBM representation of \((-1)^{v_1 v_2 v_3}\) uses two hidden neurons, as shown in \xrefname{Fig.}\cref{fig:hypergraph} (c). The method for decomposing \((-1)^{v_1v_2v_3}\) can be extended to treat \((-1)^{v_1v_2\cdots v_k}\) for arbitrary \(k\) with \(2k+O(1)\) hidden neurons \citep{S}.

\emph{RBM representation of XS/Pauli-stabilizer states.---} The Pauli stabilizer formalism generalizes graph states by applying some local Clifford transformations \citep{van2004graphical, schlingemann2001stabilizer}. The XS-stabilizer formalism generalizes the Pauli stabilizer formalism \citep{ni2015non} by changing the single qubit Pauli group to Pauli-\(S\) group \({\mathcal{P}}^S =\langle{\alpha I,X,S}\rangle\) where \(\alpha=\mathrm{e}^{\mathrm{i}\pi/4}\) and \(S=\mathrm{diag}(1,\mathrm{i})\). We have the following theorem based on the key observation that the wave function of the stabilizer state first proved in \citep{aggarwal2008boolean, dehaene2003clifford} and the similar result on XS-stabilizer state proved in \citep{ni2015non}.

\begin{theorem}
Restricted Boltzmann machines can represent any Pauli-stabilizer states and XS-stabilizer states efficiently and exactly.
\end{theorem}

Let \(\delta(x)\) satisfies \(\delta(0)=1\) and \(\delta(1)=0\). The wave functions of every Pauli-stabilizer state and XS-stabilizer state on \(n\) qubits can be written as the closed form of

\begin{align}
\label{eq:pauli}
\Psi_{\text{pauli}}(\mathbf{v})&\propto i^{l(\mathbf{v})}(-1)^{q(\mathbf{v})}\prod_j \delta(L_j^p~\text{mod }2)\\
\Psi_{\text{XS}}(\mathbf{v})&\propto {\alpha}^{l(\mathbf{v})}i^{q(\mathbf{v})}(-1)^{c(\mathbf{v})}\prod_j \delta(L_j^p~\text{mod }2),
\label{eq:pauli-xs}
\end{align}

where \(l(\mathbf{v})\), \(q(\mathbf{v})\) and \(c(\mathbf{v})\) are linear, quadratic and cubic polynomials of \(\mathbf{v}\) with integer coefficients respectively. The \(L_j^p\) are affine (linear terms plus constant term) functions of some subsets of \(\mathbf{v}\). Moreover, these polynomials \(l(x)\), \(q(x)\) and \(c(x)\) along with \(L_j^p\) can be determined efficiently from the given stabilizers.

Given this wave function, we can easily deduce its RBM representation: First, the \(\prod_j \delta(L_j^p~\text{mod }2)\) parts correspond to parity constraints, which RBM can represent. Meanwhile, functions that are products of \(i^{l(\mathbf{v})}\), \((-1)^{q(\mathbf{v})}\), \({\alpha}^{l(\mathbf{v})}\), \(i^{q(\mathbf{v})}\) and \((-1)^{c(\mathbf{v})}\) can also be represented by RBM using techniques used for graph states and hypergraph states. In the case of stabilizer states, we provide a new proof of the closed form wave function as follow: First, a stabilizer state can be generated from a Clifford circuit \citep{gottesman1997stabilizer} consisting of \(H,~S\), and \(CZ\) gates. From the encoding circuit, the stabilizer state can be represented by a DBM in the closed form Eq. \eqref{eq:pauli} directly. Then we can iteratively reduce it to an RBM while still keep the closed form \citep{S}.

We can make use of local Clifford equivalence between stabilizer state and graph states \citep{van2004graphical, elliott2008graphical} to further simplify our procedure. We can choose the encoding circuit based on fact that every stabilizer state can be chosen to be locally equivalent to a graph state such that only at most a single \(H\) or \(S\) gate acting on each qubit \citep{elliott2008graphical}. Then, the number of hidden neurons used to represent hidden variables is fewer than the number of \(H\) acting on each qubit. In total, the number of hidden neurons needed is of order \(O(N_e+n)\). The \(N_e\) is number of edges in the corresponding graph, and is at most \(O(n^2)\); however we do not usually encounter graph states from dense graphs \citep{anders2006fast}, typically only \(O(n\log n)\) or even \(O(n)\) hidden neurons are needed. Our representation method is effective and optimal. Next, we describe topological states with different topological orders within our quantum information framework. These topological states can be represented as RBM efficiently.

\emph{Fracton topological order.---} The Pauli stabilizer formalism covers most of the fracton topological order models \citep{haah2011local, yoshida2013exotic, ma2017fracton, chamon2005quantum}, such as Haah's code \citep{haah2011local}. Stabilizers of Haah's code involve two types of stabilizers on eight spins: eight \(Z\)s and eight \(X\)s in each cube.

\emph{Intrinsic topological order.---} We now consider RBM representation for some notable XS stabilizer states: the double semion (an example of string-net model \citep{levin2005string}) and many twist quantum double models. We define the double semion model on a honeycomb lattice with one qubit per edge. The wave function of double semion is: \(|\psi\rangle = \sum_{\text{$x$ is loops}}(-1)^{\text{number of loops}}|x\rangle\). In the XS-stabilizer formalism, this model has two types of stabilizer operators \citep{ni2015non}: \(g_p=\prod_{l\in v}Z_l, g_p=\prod_{l\in p}X_l \prod_{r\in \text{legs of $p$}}S_r\) corresponding to the vertex \(s\) and the face \(p\) respectively.

Quantum double models \citep{kitaev2003fault} are generalizations of the toric code that describe systems of abelian and non-abelian anyons. Twisted quantum double models further generalizes quantum double models \citep{kitaev2006anyons, buerschaper2014twisted, kitaev2003fault, kitaev2006anyons, hu2013twisted, beigi2011quantum} and are Hamiltonian realizations of Dijkgraaf-Witten topological Chern-Simons theories \citep{dijkgraaf1990topological}. Many twisted quantum double models fit into the XS-stabilizer formalism \citep{ni2015non}, thus can be represented as RBM exactly. Examples include twisted quantum double model \(D^{\omega}(\mathbb{Z}_2^n)\) with the group \(\mathbb{Z}_2^n\) and different twists \(\omega\in H^3(\mathbb{Z}^n,U(1))\) on a triangular lattice, where \(H^3(\mathbb{Z}_2^n, U(1))\) is the third cohomology group. The \(H^3(\mathbb{Z}_2^1, U(1))\) case includes the double semion. It is known \citep{ni2015non, de1997spontaneously} that a non-trivial twist from \(H^3(\mathbb{Z}_2^3, U(1))\) harbors non-abelian anyon excitations.

\begin{figure}
\centering
\includegraphics[width=0.40000\textwidth]{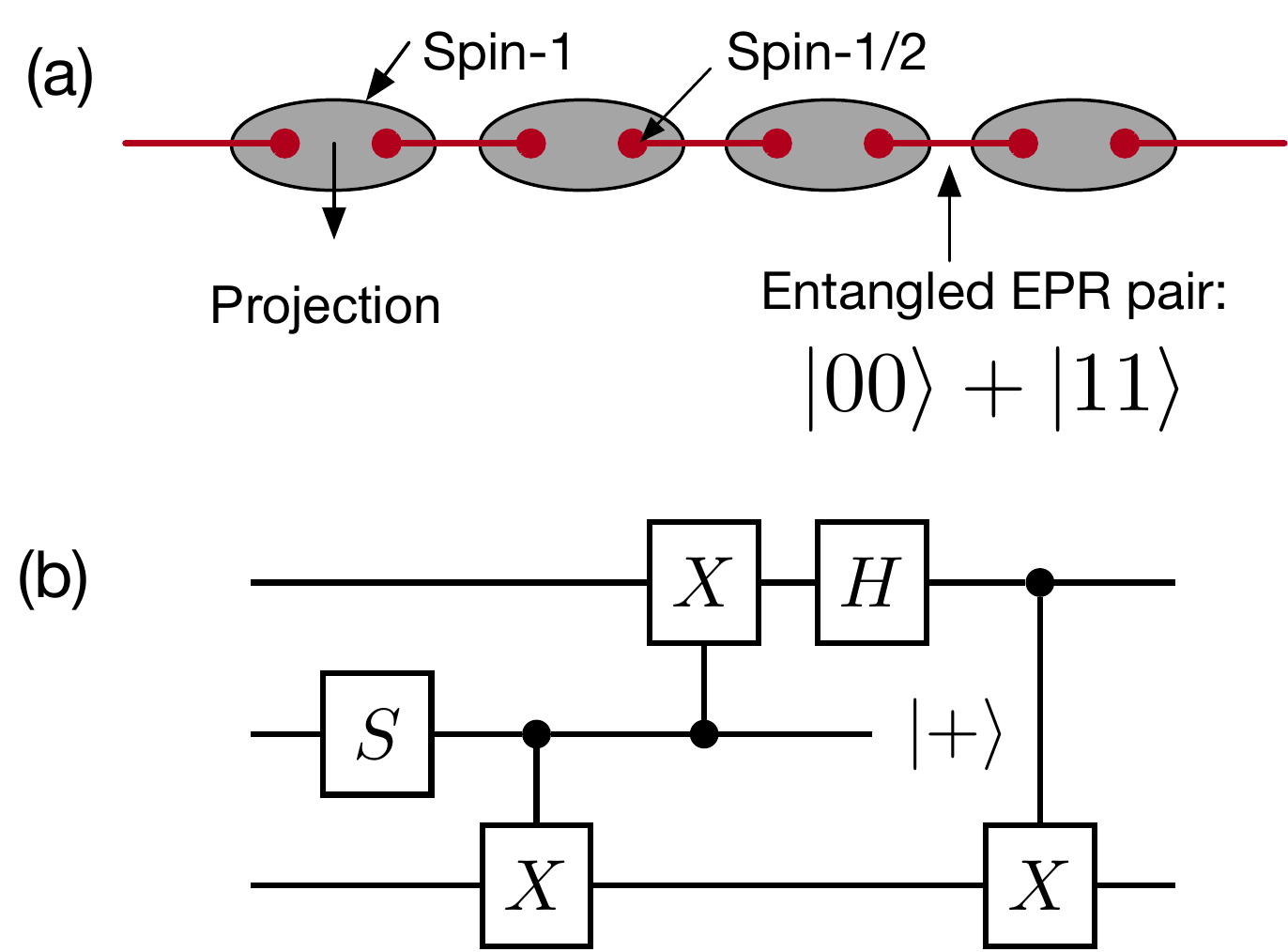}
\caption{(a) The ground state of the AKLT model. The spin one on each lattice site can be decomposed into two spin 1/2's which form EPR pairs between nearest neighbor pairs. At the two ends of an open chain, there are two isolated spin 1/2's giving rise to a four-fold ground state degeneracy. Every red line represent a \(|00\rangle+|11\rangle\) state. Each shaded circle represent a projection operator \(P=|100\rangle(\langle01|+\langle 01|)+|010\rangle i(\langle 01|-\langle 10|)+|001\rangle(\langle 00|-\langle 11|)\). (b) The Clifford circuit that realizes the projection \(P\) under the unary representation (living in the space spanned by \(|100\rangle, ~|010\rangle\) and \(|001\rangle)\). Here \(|+\rangle\) means projecting onto \(|+\rangle\) state.\label{fig:AKLT}}
\end{figure}

\emph{Symmetry protected topological order.---} The AKLT model \citep{affleck1988valence, affleck1987rigorous} is a one-dimensional spin-1 model with symmetric protected topological order. When imposing the periodic boundary condition, the unique ground state \(|\psi_{AKLT}\rangle\), in terms of the matrix product state, writes
\begin{equation}
\Psi_{AKLT}(a_1,a_2,\cdots,a_n)\propto\text{Tr}(A_{a_1}A_{a_2}\cdots A_{a_n}),
\label{eq:AKLT}\end{equation}
where \(A_{-1}=X,~A_{0}=Y\), \(A_{+1}=Z\) and \(a_i=-1,0,+1\). Alternatively, matrix product states can be described as projected entangled pairs states (PEPS) \citep{zeng2015quantum}. As shown in \xrefname{Fig.}\cref{fig:AKLT}, every red line is a EPR pair \(|00\rangle+|11\rangle\). Each shaded circle represents a projection from two spins of dimension 2 to a physical degree of dimension \(3\) (spin-1). \(P=\sum_{a_i,\alpha,\beta} A_{a_i,\alpha,\beta}|a_i\rangle\langle \alpha\beta|\) where the summation is over \(a_i=-1,0,1\) and \(\alpha,\beta=1,2\). After using unary representation, in the quantum circuit language, the projection \(P\) is a map that maps \(|01\rangle+|10\rangle \to |100\rangle,~i(|01\rangle-|10\rangle)\to |010\rangle\) and \(|00\rangle-|11\rangle\to |001\rangle\). We find such a quantum circuit made of Clifford gate, as shown in \xrefname{Fig.}\cref{fig:AKLT}(b). Because all operations are Clifford gates, the whole quantum state is a Pauli stabilizer state restricted to the single excitation subspace. Thus it can be represented by RBM with unary hidden neurons (blue neurons in \xrefname{Fig.}\cref{fig:unary}) and other hidden neurons \citep{S}.

We also show RBM can represent other symmetry-protected topologically ordered states. These examples include the 2D CZX model (with \(\mathbb{Z}_2\) symmetry), and Yoshida's CCZ model \citep{yoshida2016topological} (with \(\mathbb{Z}_2\otimes \mathbb{Z}_2\otimes \mathbb{Z}_2\) symmetry). The ground state of the 2D CZX model \citep{chen2011two} is a tensor product of GHZ states \(|0000\rangle+|1111\rangle\) on each plaquette. Because GHZ state belongs to stabilizer states, RBM can exactly represent the ground state of the 2D CZX model. Similar to SPT cluster states, the ground state of Yoshida's CCZ model \citep{yoshida2013exotic} on the trivalent lattice is a hypergraph state with 3-hyperedges. Thus it can also be represented efficiently by RBMs.

\emph{Discussions.---} This paper sets out to investigate different topological models with RBM representation using tools from quantum information such as the (XS) stabilizer and the (hyper) graph-state formalisms. The most significant findings are the fact that RBM states can capture ground states of exotic models with different types of topological orders, including intrinsic topological orders, symmetry protected topological orders and fracton topological orders. It remains open whether RBM can capture additional string-net models with more exotic non-abelian anyon excitations such as the states of the double Fibonacci model \citep{levin2005string}. RBM state may be helpful in investigating symmetry enriched topological order \citep{cheng2017exactly} and symmetry fractionalization \citep{chen2017symmetry}. Our exact representation results provide useful guidance for future numerical studies on topological phase transitions.

Our results are also of interest from the quantum information perspective. The Gottesman-Knill theorem \citep{gottesman1998heisenberg, anders2006fast, aaronson2004improved} states that stabilizer dynamics can be efficiently simulated. Our result which shows RBM states contain stabilizer states suggests classical simulation of an unknown larger family of quantum circuits may benefit from the RBM representation \citep{jonsson2018neural}. Because hypergraph states allow for an exponentially increasing violation of the Bell inequalities \citep{gachechiladze2016extreme, guhne2014entanglement, bell2001einstein, RevModPhys.86.419}, our results provide analytical evidence for numerical studies \citep{deng2017machine} on estimating maximum violation of Bell inequalities.

\begin{acknowledgments}
We thank Giuseppe Carleo, Dong-Ling Deng, Sheng-Tao Wang, Mingji Xia, Shunyu Yao, Fan Ye, Zhengyu Zhang and You Zhou for helpful discussions.  This work was supported by Tsinghua University and the National Key Research and Development Program of China (2016YFA0301902). 

\textit{Note added.—} After completion of this work, a preprint appeared in arXiv \citep{zhang2018efficient} which gives a different method to represent stabilizer states with the restricted Boltzmann machine.
\end{acknowledgments}

\nocite{cai2014complexity,cai2018clifford}


\begin{thebibliography}{75}
\bibitem [{aff()}]{affine}%
  \BibitemOpen
  \href@noop {} {\ }\bibinfo {note} {This can be generalized to affine
  function: both equals to 0 and 1 are allowed. The RBM representation of toric
  code uses the case of $k=4$.}\BibitemShut {Stop}%
\end{thebibliography}

%

\appendix
\pagebreak
\widetext

\section{Detailed derivation of RBM representations of hypergraph states}\label{detailed-derivation-of-rbm-representations-of-hypergraph-states}

First we recall the definition of quantum hypergraph state in more detail. Mathematically, hypergraphs are generalization of graphs in which an edge may connect more than two vertices. Formally, a hypergraph \(H\) is a pair \(H=(X,E)\) where \(X\) is a set of element called vertices, and \(E\) is a subset of \(P(X)\), where \(P(X)\) is the power set of \(X\). Given a mathematical hypergraph, the corresponding quantum state can be generated by following similar steps in constructing a graph state: first, assign to each vertex a qubit and initialize each qubit as \(|+\rangle\); Then, whenever there is hyperedge, perform a controlled-\(Z\) operation between all connected qubits; if the qubits \(v_1,v_2,\dots,v_k\) are connected by a \(k\)-hyperedge, then perform \(C^{k}Z_{i_1,i_2,\dots,i_k}\). As a result, the hypergraph state and its wave function \citep{rossi2013quantum} take the form
\begin{equation}
\begin{aligned}
|g\rangle &= \prod_{\{i_1,i_2,\dots,i_k\}\in E}C^k Z_{i_1,i_2,\dots,i_k}|+\rangle^{\otimes n}\\
\Psi(g) &=\sum_{v_1,v_2,\cdots,v_n}\prod_{\{i_1,i_2,\dots,i_k\}\in E}(-1)^{i_1i_2\dots i_k}|v_1 v_2\cdots v_n\rangle.
\end{aligned}
\label{eq:hypergraph-CCZ}\end{equation}
Next, we give a detailed derivation of realizing correlation factor of the type \((-1)^{v_1v_2\cdots v_k}\) using restricted Boltzmann machines. We first recall \((-1)^{v_1v_2}\) which relates to graph states from \citep{Gao:2017uk}:
\begin{equation}
\begin{aligned}
H_{v_1,v_2}&=\frac{(-1)^{v_1v_2}}{\sqrt{2}}=\sum_{h=0,1}e^{W_H(v_1,h)+W_H(v_2,h)}\\
&=\cos\left(\frac{\pi}{4}[2(v_1+v_2)-1]\right).
\end{aligned}
\label{eq:Hv1v2}\end{equation}
Note the above equation \xrefname{Eq.}\cref{eq:Hv1v2} is only true for \(v_i=0,1\). By \(\cos v=\frac{1}{2}(e^{iv}+e^{-iv})=\sum_{h} e^{iv(2h-1)-\ln 2}\), we have an explicitly formula fo \(W_H\):
\[
\begin{aligned}
W_H(v_i,h)&=i\pi v_i h-i\pi[2v_i+h]/4+(i\pi/4-\ln 2)/2\\
&=\frac{\pi}{8}i-\frac{\ln 2}{2}-\frac{\pi}{2}iv-\frac{\pi}{4}ih+i\pi vh.
\end{aligned}
\]
Then we consider \((-1)^{v_1v_2v_3}\) with the following decomposition.
\[
\begin{aligned}
&(-1)^{v_1v_2v_3}\\
&=(-1)^{v_1+v_2+v_3}\cdot\left[\frac{8}{3}\cos^2\left(\frac{2\pi}{3}(v_1+v_2+v_3-1)+\frac{\pi}{2}\right)-1\right]\\
&=e^{i\pi(v_1+v_2+v_3)}\cdot\left[\frac{1}{3}+\frac{4}{3}\cos\left[\frac{4\pi}{3}(v_1+v_2+v_3)-\frac{\pi}{3}\right]\right].
\end{aligned}
\]
Let \(v=\frac{4\pi}{3}(v_1+v_2+v_3)-\frac{\pi}{3}\). The above equation simplifies to
\[
\begin{aligned}
(-1)^{v_1v_2v_3}&=e^{i\pi(v_1+v_2+v_3)}\cdot\left(\frac{1}{3}+\frac{4}{3}\cos v\right)\\
&=e^{i\pi(v_1+v_2+v_3)}\cdot\left(\frac{1}{3}+\frac{2}{3}(e^{iv}+e^{-iv})\right).
\end{aligned}
\]
The RBM can represent first part by definition. We consider the second part as a RBM with two hidden neurons and set equations:
\[
\begin{aligned}
&\frac{1}{3}+\frac{2}{3}(e^{iv}+e^{-iv})=\sum_{h_1,h_2} e^{wvh_1+w_2vh_2+b_1h_1+b_2h_2+c}\\
&=e^c(1+e^{w_1v+b_1}+e^{w_2v+b_2}+e^{(w_1+w_2)v+b_1+b_2}).
\end{aligned}
\]
Solving the above equation, we get \(w_1=-w_2=\sqrt{-1}\) , \(b_1=b_2=b\), and
\[
\begin{aligned}
&e^c(1+e^{2b})=\frac{1}{3},\\
&e^{b+c}=\frac{2}{3}.
\end{aligned}.
\]
Solve the above system of equations, we obtain
\[
\begin{aligned}
&b=\ln(\frac{1\pm i\sqrt{15}}{4}),\\
&c=\ln\frac{2}{3}-b.
\end{aligned}
\]
In fact, this construction can be extended to any function of the form \((-1)^{v_1v_2\cdots v_k}\). The proof goes as follows. First, we generlize our construction a bit to simulate the correlation factor \(2A\cos(\sum_{i=1}^k v_i)+B\) by two hidden neurons. Setting the equation:
\[
\sum_{h_1,h_2} e^{w_1(\sum_{j=1}^k v_j)h_1+w_2(\sum_{j=1}^k v_j)h_2+b_1h_1+b_2h_2}=A(e^{i(\sum_{i=1}^k v_j)}+e^{-i(\sum_{j=1}^k v_j)})+B.
\]
Simplifying above equation, we obtain
\[
\begin{aligned}
&w_1=-w_2=i\\
&e^{c+b_1}=e^{c+b_2}=A,\\
&e^c(1+e^{b_1+b_2})=B.
\end{aligned}
\]
Solving the system of equation, we get
\[
\begin{aligned}
&b=b_1=b_2=\ln (\frac{B\pm \sqrt{B^2-4A^2}}{2A}),\\
&c=\ln A-b=\ln A=\ln \frac{2A^2}{B\pm \sqrt{B^2-4A^2}}.
\end{aligned}
\]
Now we proceed to construct RBM representation for \(g(v_1,v_2,\cdots,v_k)=(-1)^{v_1 v_2\cdots v_k}\). Note \(g(v_1,v_2,\cdots,v_k)\) equals to 1 only if \(v_1=v_2=\cdots v_k=1\), i.e., \(v_1+v_2+\cdots+v_k=k\).For convenience, we first consider
\[
f(\sum_{i=1}^k v_i)=\frac{1}{2}(1-g(\sum_{i=1}^k v_i))=\begin{cases} 
1 & v_1+v_2+\cdots v_k=k \\
0 & \text{otherwise}
\end{cases}
\]
The trick of the construction is to introduce the function
\[
t(\sum_{i=1}^k v_i)=\cos\left(\frac{2\pi}{k+1}(v_1+v_2+\cdots +v_k+1)\right)=\begin{cases} 
1 & v_1+v_2+\cdots v_k=k \\
\text{other values} & \text{otherwise}
\end{cases}
\]
Then the function \(f\) can be chosen as (the idea is similar to Lagrange polynomial method)
\[
f(\sum_{i=1}^k v_i)=\prod_{i=0}^k \frac{t(\sum_{i=1}^k v_i)-t(i)}{t(i)-t(1)}=\begin{cases} 
1 & v_1+v_2+\cdots v_k=k \\
0 & \text{otherwise}
\end{cases}
\]
Then by substitution we get \(g\)
\[
g=1-2f=1-2\prod_{i=0}^k \frac{t(\sum_{i=1}^k v_i)-t(i)}{t(i)-t(1)}
\]
In principle we can factorize \(g\) into the form of \(g(t)=\prod_i(2C_i t+D_i)\), which can then be simulated by RBM term by term.

Now we remark on the relation between our decomposition and Eq. (22) in \citep{carleo2018constructing}. These two mainly differs from in the choose of the basis. We use \(\{0,1\}\) basis in contrast to the \(\{-1,+1\}\) basis in \citep{carleo2018constructing}. This difference leads to several consequences. First, in the \(\{0,1\}\) basis, \((-1)^{x_1 x_2 x_3}\) only equals to \(-1\) when all the \(x_i\) equals to 1. While in the \(\{-1,+1\}\) basis, \({e}^{x_1 x_2 x_3V}\) equals to \(e^V\) or \(e^{-V}\) if number of -1 in \(x_i\) is even/odd. That's the reason why latter decomposition only needs two hidden neurons. If we derive a similar RBM expression of our n-body gadget from Eq. (22) in \citep{carleo2018constructing} directly, \(O(2^k)\) hidden neurons is needed. In contrast our cost is \(2k+O(1)\). Meanwhile, the correlation factor of graph/hypergraph states \((-1)^{x_1 x_2 x_3}\) is somewhat meaningless in the \(\{-1,+1\}\) basis.

\section{Derivation details of RBM unary representations}\label{derivation-details-of-rbm-unary-representations}

Restricting three neurons only take values from 100, 010 and 001 is equvalently to construct to RBM representation of \(W\) state: \(|W\rangle =( |001\rangle+|010\rangle+|001\rangle)/\sqrt{3}\). This can be done by two hidden neurons as follow. Consider the function
\[
h(v_1,v_2,v_3)=\begin{cases} 
1 & v_1+v_2+v_3=1 \\
0 & \text{otherwise}
\end{cases}
\]
We find a decomposition of \(h(v_1,v_2,v_3)\):
\[
h(v_1,v_2,v_3)=(-1)^{v_1+v_2+v_3}\times\left(-\frac{1}{3}+\frac{2}{3}\cos\left(\frac{4\pi}{3}(v_1+v_2+v_3)-\frac{\pi}{3}\right)\right),
\]
which can be simulated using two hidden neurons using the method from the last section. The weight can be computed.

The notion of \(W\) state has been generalized for \(n\) qubits to Dicke state. We define \(W_{n,k}\) state to be a uniform superposition of all computational basis states \(|x\rangle\) where \(x\) is a Hamming weight \(k\) bit string. It can be done by using nearly the same technique from the last section.

\section{Derivation details on RBM representation of stabilizer states}\label{derivation-details-on-rbm-representation-of-stabilizer-states}

\begin{figure}
\centering
\includegraphics[width=0.45000\textwidth]{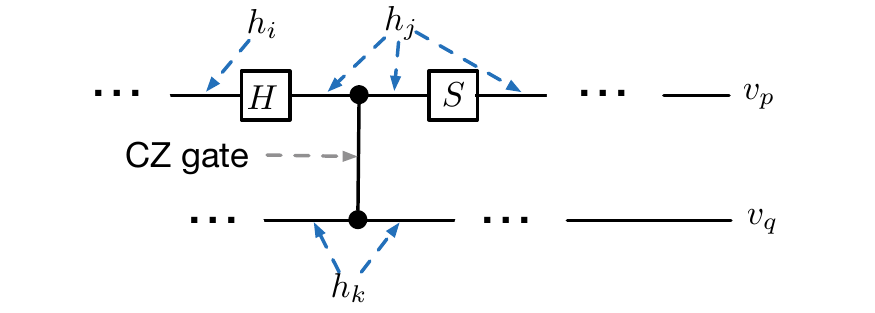}
\caption{The close-up of a Clifford circuit. Visible neurons are \(v_p,v_q,\cdots\) and hidden neurons are \(h_i,h_j,\cdots\). Each \(H\) gate corresponds to two neurons with the weight \((-1)^{x_ix_j}\); Each \(S\) gates corresponds to one neuron with the weight \(i^{x_j}\); Each \(CZ\) gate corresponds to two neurons with the weight \((-1)^{x_1x_2}\). The whole wave function is a DBM which can be eliminated to a RBM.\label{fig:clifford}}
\end{figure}

The wave function can be represented by a DBM as follow. We denote visible neurons as \(v_p,v_q,\cdots\) and hidden neurons as \(h_i,h_j,h_k\cdots\). For convenience. We unify the notation of these neurons as \(x_i,x_j,x_k,\cdots\) here. As shown in \xrefname{Fig.}\cref{fig:clifford}: each \(H\) gate corresponds to two neurons with the weight \((-1)^{x_ix_j}\); each \(S\) gates corresponds to one neuron with the weight \(i^{x_j}\) (even though each single qubit matrix has two indices); each \(CZ\) gate corresponds to two neurons with the weight \((-1)^{x_1x_2}\). Multiplying together and sum over all the hidden indices, we can get the wave function of the stabilizer state:
\begin{equation}
\Psi(v_p,v_q,\cdots)\propto\sum_{h_i,h_j,\cdots}i^{L(h_i,h_j,\cdots,v_p,v_q,\cdots)}(-1)^{Q(h_i,h_j\cdots,v_p,v_q,\cdots)},
\label{eq:closed_form}\end{equation}
where \(L\) and \(Q\) are affine (with linear and constant terms) and quadratic (without linear term) polynomials with integer coefficients.

Next, we reduce this DBM to a RBM with the following strategy: after eliminating a \(h_i\) for some \(i\), the remaining term still keeps the form shown in \xrefname{Eq.}\cref{eq:closed_form} except an additional parity constraint on some of \(v_p,v_q,\cdots\). Here, we consider the effect of eliminating \(h_i\) in detail:

\begin{enumerate}
\def\labelenumi{(\arabic{enumi})}
\tightlist
\item
  If the coefficient of \(h_i\) in \(L\) is 0 or 2, ignoring the terms not depending on \(h_i\), the remaining in \xrefname{Eq.}\cref{eq:closed_form} could be written as:
  \[
  \sum_{h_i}(-1)^{h_iL_i^\prime}=2\delta(L_i^p\mod2)
  \]
  where \(\delta\) means constraint and \(L_i^p\) is an affine polynomial involves those \(x_j\) which interact with \(h_i\) in \(Q\) and constant terms which are the coefficients of \(h_i\) in \(L\). There are two cases for \(L_i^p\):
\end{enumerate}

(1.1) If \(L_i^p\) only involves visible neurons, we can put this constraint \(\delta\) before the summation in \xrefname{Eq.}\cref{eq:closed_form};

(1.2) If \(L_i^p\) involves hidden neurons, e.g. \(h_j\), then we can solve the equation \(L^p_i=0\mod 2\) by \(h_j=L^{p\prime}_i\mod 2\). After the summation of \(h_j\), the only remaining terms are those satisfying \(h_j=L^{p\prime}_i\mod 2\) so we can replace \(h_j\) by \(L^{p\prime}_i\) in \(Q\) and by \((L^{p\prime})^2\) in \(L\) (because \(L^{p\prime}\mod2=(L^{p\prime})^2\mod4\)) in \xrefname{Eq.}\cref{eq:closed_form}. The key point is: the coefficients of quadratic terms in \((L^{p\prime})^2\) are always 2 thus they keeps the same form as the terms in the summation of \xrefname{Eq.}\cref{eq:closed_form} after summation over \(h_i\) and \(h_j\).

\begin{enumerate}
\def\labelenumi{(\arabic{enumi})}
\setcounter{enumi}{1}
\tightlist
\item
  If the coefficient of \(h_i\) in \(L\) is 1 (the case for 3 is similar), ignoring the terms not depending on \(h_i\), the remaining in \xrefname{Eq.}\cref{eq:closed_form} can be written as:
  \[
  \sum_{h_i}(-1)^{h_iL_i^p}i^{h_i}=(1+i)i^{3(L_i^{p})^2}
  \]
  The key point is that the coefficient of quadratic terms in \((L_i^{p})^2\) is 2 thus it keeps the same form as the terms in the summation of \xrefname{Eq.}\cref{eq:closed_form} after summation over \(h_i\)
\end{enumerate}

After eliminating all the hidden variables, the wave function could be written as the form of Eq. (3) . The parity constraints could be represented by a hidden neuron with the corresponding visible neurons as shown before.

Our method also generalizes conveniently to qudit stabilizer states \citep{hostens2005stabilizer} with the proposed unary representation.

\end{document}